# `maplet`: An extensible R toolbox for modular and reproducible omics pipelines

Kelsey Chetnik[1], Elisa Benedetti[1], Daniel P. Gomari[2], Annalise Schweickart[1], Richa Batra[1], Mustafa Buyukozkan[1], Zeyu Wang[1], Matthias Arnold[2], Jonas Zierer[1,3], Karsten Suhre[4], Jan Krumsiek[1,#]

[1] Department of Physiology and Biophysics, Institute for Computational Biomedicine, Englander Institute for Precision Medicine, Weill Cornell Medicine, New York, NY 10021, USA

[2] Institute of Computational Biology, Helmholtz Zentrum München—German Research Center for Environmental Health, 85764 Neuherberg, Germany

[3] current address: Novartis Institutes for Biomedical Research (NIBR), Novartis, Basel, Switzerland

[4] Department of Physiology and Biophysics, Weill Cornell Medical College—Qatar Education City, Doha, Qatar

[#] corresponding author: Jan Krumsiek, jak2043@med.cornell.edu

## Abstract

This paper presents `maplet`, an open-source R package for the creation of highly customizable, fully reproducible statistical pipelines for omics data analysis, with a special focus on metabolomics-based methods. It builds on the `SummarizedExperiment` data structure to create a centralized pipeline framework for storing data, analysis steps, results, and visualizations. `maplet`'s key design feature is its modularity, which offers several advantages, such as ensuring code quality through the individual maintenance of functions and promoting collaborative development by removing technical barriers to code contribution. With over 90 functions, the package includes a wide range of functionalities, covering many widely used statistical approaches and data visualization techniques.

**Availability and implementation**

The maplet package is implemented in R and freely available at https://github.com/krumsieklab/maplet.

# 1 Introduction

A major shift within the biomedical community in recent years has been a push to promote reproducibility in research[1–3]. This has led to substantial changes in scientific publishing, including new rules for the mandatory sharing of source code and accompanying data for publication in peer-reviewed journals[4]. Adapting data analysis workflows to these new prerequisites often requires significant time and effort to develop code that is easily readable, robust, and sharable.

The most effective way to facilitate reproducibility in everyday data analysis workflows is through *modular* toolboxes that automate large parts of a typical omics data processing pipeline while also allowing for high customizability. A modular toolbox is designed such that specific tasks are encapsulated as individual functions. This design offers several advantages, such as ensuring robust code through the individual development, testing, and maintenance of functions and promoting collaborative development by removing technical barriers to code contribution.

Here we present `maplet`, an open-source R package that provides a modular pipeline framework for flexible and reproducible omics data analysis, with a special focus on metabolomics data. `maplet` contains a diverse collection of functions, covering preprocessing, statistical analysis, pathway analysis, visualization, and various other functionalities. The toolbox is under active development by an international team. While there are other toolboxes for creating analytical pipelines, such as `MetaboAnalystR`[6] or `structToolbox`[7], none provide the same degree of reproducibility and extensibility offered by `maplet`.

# 2 Toolbox

## 2.1 Pipeline Design

The `maplet` package allows for the creation of fully reproducible analytical pipelines. This is achieved by using a centralized pipeline object that is passed between each function and records all data, results, plots, analysis steps and their parameters. The pipeline object builds on `SummarizedExperiment`[6], a container class provided by Bioconductor[7] which stores datasets and all corresponding annotations in a single object.

`maplet` is designed to be used with a pipe operator – either the popular `%>%` operator from the `magrittr` package[8] or the recently introduced `|>` operator from base R. Pipe operators enable the smooth connection of processing steps in a `maplet` pipeline - seamlessly passing the container object from function to function. This makes code more readable and eliminates the need for intermediate result variables. Figure 1 presents a subsection of a pipeline and a diagram representing how each step in the pipeline is stored in the container object.

## 2.2 Modularity

`maplet` follows a modular 'one function, one operation' design. Each task is encapsulated in a single function, which enables the rapid development of pipelines where any step can be flexibly inserted, removed, or rearranged. Another key advantage of this modular design is the ability to maintain high-quality code. Since functions have no interdependencies, they can be rigorously evaluated and maintained separately. Finally, the modular structure promotes a culture of open-source development by removing the technical barriers to code contribution for unfamiliar developers. Any interested user can add a desired functionality based on a simple function template and only minimal knowledge of the inner workings of the package.

## 2.3 Functionality

Currently, `maplet` contains a growing set of over 90 functions organized into various groups, such as data loading, annotation, data modification, preprocessing, statistical analysis, visualization, reporting results, exporting data, and pipeline maintenance. This covers many commonly used analytical methods necessary for standard data analysis encountered in everyday research projects. There are specialized loading functions for working with data from various popular omics platforms and a wide variety of functionalities commonly used by bioinformatics researchers, including linear models, missing-value imputation, PCA, heatmaps, as well as more advanced functionalities such as pathway analysis and network inference. The `maplet` package comes with several extensive example pipelines and documentation to aid new users in the design of new workflows.

## 2.4 Report Generation and Result Access

Once a `maplet` pipeline has been executed, results can be visualized through comprehensive reports automatically assembled by `maplet` using R `markdown/knitr`. These reports lay out all functions in the pipeline in the order they were executed, including the name of the function, arguments, and any plots or statistics tables produced by the function. The report is compiled into a single HTML, PDF, or Word document, which stores all results in a single location and can be easily shared. Moreover, `maplet` comes with a series of accessor functions, which allow the user to extract processed data, statistical results or plots from the pipeline object and further analyze them using their own R code.

# 3 Conclusion

The `maplet` R package facilitates the fast development of reproducible analysis pipelines for omics data. Its modular design allows for highly customizable, fully reproducible omics pipelines, while also improving readability, ensuring code quality, and promoting open-source development.

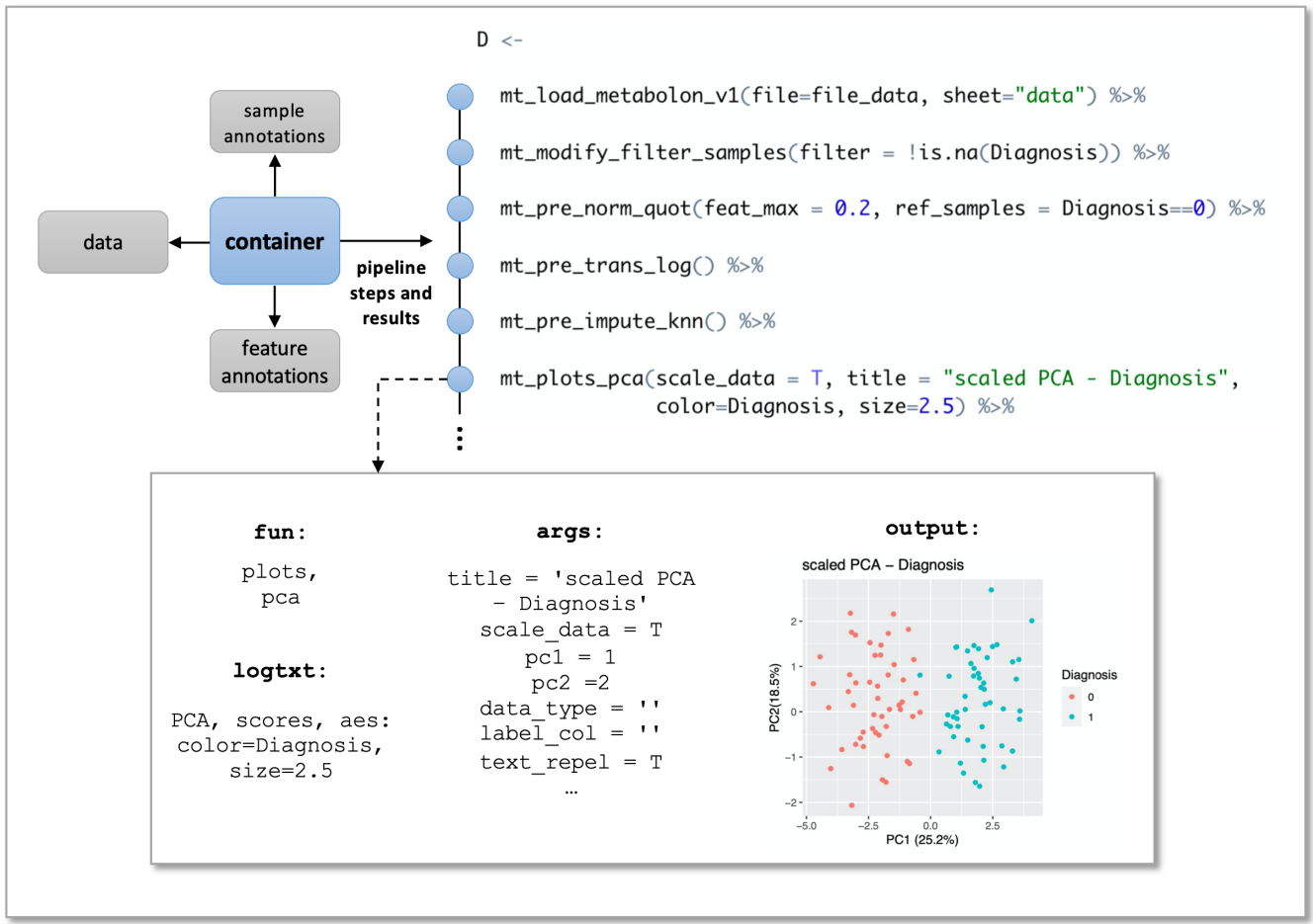

**Figure 1.** `maplet` pipeline. All data and annotations are stored in a central `SummarizedExperiment` object, which is passed between functions. Each function generates a result entry, containing all function-specific information as well as the results the function generated (such as statistics tables and plots).

## Acknowledgements & Funding

KS is supported by 'Biomedical Research Program' funds at Weill Cornell Medical College in Qatar, a program funded by the Qatar Foundation and multiple grants from the Qatar National Research Fund (QNRF). JK is supported by the National Institute of Aging of the National Institutes of Health under award 1U19AG063744.